# Tunable light-emission through the range 1.8–3.2 eV and p-type conductivity at room temperature for nitride semiconductors, Ca(Mg$_{1-x}$Zn$_x$)$_2$N$_2$ ($x = 0 - 1$)

Masatake Tsuji,[1] Hidenori Hiramatsu,[1,2,a] and Hideo Hosono[1,2]

1: Laboratory for Materials and Structures, Institute of Innovative Research, Tokyo Institute of Technology, Mailbox R3-3, 4259 Nagatsuta-cho, Midori-ku, Yokohama 226-8503, Japan

2: Materials Research Center for Element Strategy, Tokyo Institute of Technology, Mailbox SE-1, 4259 Nagatsuta-cho, Midori-ku, Yokohama 226-8503, Japan

a) Electronic mail: h-hirama@mces.titech.ac.jp

Abstract

The ternary nitride $CaZn_2N_2$, composed only of earth-abundant elements, is a novel semiconductor with a band gap of ~1.8 eV. First-principles calculations predict that continuous Mg substitution at the Zn site will change the optical band gap in a wide range from ~3.3 eV to ~1.9 eV for $Ca(Mg_{1-x}Zn_x)_2N_2$ ($x$ = 0–1). In this study, we demonstrate that a solid-state reaction at ambient pressure and a high-pressure synthesis at 5 GPa produce $x$ = 0 and 0.12, and 0.12 < $x$ ≤ 1 polycrystalline samples, respectively. It is experimentally confirmed that the optical band gap can be continuously tuned from ~3.2 eV to ~1.8 eV, a range very close to that predicted by theory. Band-to-band photoluminescence is observed at room temperature in the ultraviolet–red region depending on $x$. A 2% Na doping at the Ca site of $CaZn_2N_2$ converts its highly resistive state to a p-type conducting state. Particularly, the $x$ = 0.50 sample exhibits intense green emission with a peak at 2.45 eV (506 nm) without any other emission from deep-level defects. These features meet the demands of the III-V group nitride and arsenide/phosphide light-emitting semiconductors.

TOC Figure

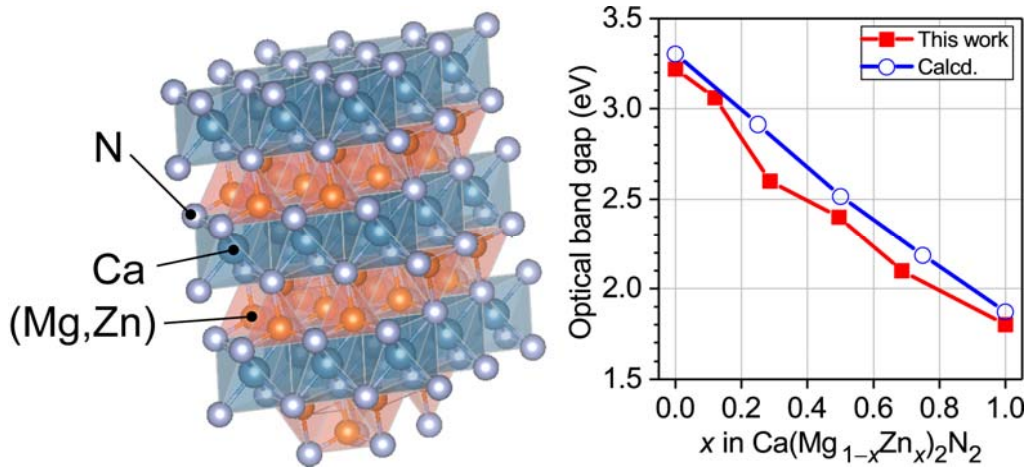

Rapid exploration of novel functional materials, composed only of earth-abundant elements, has urgently been required for the future realization of a sustainable society. Furthermore, such functional materials will lead to highly efficient energy conversion, storage, and optoelectronic functionalities, such as electron transport and light emission, and will be able to replace those materials currently being used, such as heavy, rare, and toxic elements. To fulfill such a role, nitrides are promising candidates because they are environmentally benign and remain unexplored compared with other materials systems, such as oxides, chalcogenides, and pnictides (e.g., arsenides and phosphides). The primary reason for such exploration results from the very high stability of a nitrogen molecule with triple bonding compared with an oxygen molecule, chalcogen, or pnictogen. $N_2$ is therefore inactive even at high temperatures and under high pressures without appropriate catalysts, which has led to the historical reluctance for the wider exploration of nitrides compared with other compounds. This is the reason associated with the requirements of more active nitrogen sources, such as ammonia and discharged plasma, for novel nitride research, which themselves involve an immense amount of time and effort to obtain. Thus, the currently commercialized nitride semiconductors have mostly been limited to III-V group nitrides, GaN and its based alloys, with appropriate band gaps intentionally tuned by solid-solution alloying techniques.

A light-emitting semiconductor with electron current injection is one of the key components in practical optoelectronic devices. Although such components have been commercially used, especially the III-V group alloy semiconductors, such as (Ga,Al,In)N- and (Ga,Al,In)(As,P)-based materials, there remains a strong demand on next-generation materials to exhibit a higher performance in terms of light brightness, quantum efficiency, and color accuracy (i.e., emission bandwidth) [1]. However, irrespective of the rapid growth in science and technology, light-emitting semiconductors have a serious issue, so called the green gap problem, where the emission quantum efficiency drastically decreases to ca. 20% in the particular wavelength region around green light, even for the practical III-V group-based light-emitting diodes [2]. This is mainly because of the large in-plane lattice mismatch between In-based III-V alloys and single-crystal substrates, and the thermal instability of In-based III-V nitride alloys. Thus, no highly efficient green-light emitting semiconductor material has yet been realized for next-generation optoelectronic devices. Therefore, it is necessary to explore and discover novel semiconductor materials, which possibly have higher thermal stability and more compatible in-plane lattice mismatches with III-V group nitrides, with a high quantum efficiency of light emission in the green wavelength region (band gap = 2.3−2.5 eV, wavelength $\lambda$ = 500−550 nm).

Recent progress of materials informatics, which combines comprehensive first-principles calculations with experimental validations, has led to an acceleration in the discoveries and/or predictions of novel ternary nitride semiconductors, such as $ZnSnN_2$ [3], $Cu(Nb, Ta)N_2$ [4], $LaWN_3$ [5], and $CaZn_2N_2$ [6], which all were previously unknown. Therefore, materials informatics provides the possibility to overcome the above-mentioned historical lack of exploration research of novel functional nitride compounds. $CaZn_2N_2$ [6] was predicted through such an approach, and was proposed with experimental validation as a novel promising optoelectronic functional nitride semiconductor with a band gap of ~1.8 eV because of its small electron and hole effective masses and direct transition-type energy band structure with a wide tunability of its optical band gap of isomorphism nitrides by an alloying technique similar to III-V group semiconductors. In Ref. [6], it was predicted that the band gaps could be widely tuned from ~3.3 eV (for $CaMg_2N_2$, i.e., ultraviolet region) to ~1.6 eV (for $SrZn_2N_2$) and ~0.4 eV (for $CaCd_2N_2$) (i.e., red–infrared region) by solid-solution alloying at both the Ca and Zn sites. This prediction indicates that this nitride is a promising host candidate as a light-emitting semiconductor with high electron and hole transport properties in the widely tunable ultraviolet–infrared wavelength region.

In this study, we focused on an earth-abundant $Ca(Mg_{1−x}Zn_x)_2N_2$ ($x$ = 0–1) complete solid-solution nitride system because its band gap is controllable in the range of ~1.9 eV to ~3.3 eV, following the computational prediction. Here we experimentally report the validity of the prediction; that is, its band gap is tunable, the $x$ = 0.50 sample exhibits an intense green band-to-band emission without deep defect emission even at room temperature and it can be doped to a p-type semiconductor by 2% Na doping at the Ca site.

**Structure and phase purity of Ca(Mg$_{1-x}$Zn$_x$)$_2$N$_2$ ($x$ = 0–1)**

Figure 1(a) shows XRD patterns of polycrystalline Ca(Mg$_{1-x}$Zn$_x$)$_2$N$_2$ ($x$ = 0–1). The results of Rietveld analysis are summarized in Fig. S1. We confirmed that the main crystalline phase of all the samples had the same crystal structure as that of both end members, CaMg$_2$N$_2$ [7] and CaZn$_2$N$_2$ [6], which consisted of a hexagonal structure of a trigonal system with $P\bar{3}m1$ (No. 164); that is, successful fabrication of complete solid-solution samples between $x$ = 0 and 1. With an increase in $x$, the three diffraction peaks of $10\bar{1}1$ & $01\bar{1}1$, $10\bar{1}2$ & $01\bar{1}2$, and $0003$ of the Ca(Mg$_{1-x}$Zn$_x$)$_2$N$_2$ main phase, which are surrounded by dotted squares in Fig. 1(a), were continuously shifted to higher $2\theta$ positions. Corresponding to this peak shift, both the $a$- and $c$-axis lattice parameters shrank with increasing $x$ [see Fig. 1(b)]. Compared with the calculated results by first-principles structure relaxation [6], the absolute values of both end members ($x$ = 0 and 1) and their variation trend with a change in $x$ well-agreed with each other, although the calculated values were slightly smaller than the observed values. Because the ionic radii of Mg$^{2+}$ and Zn$^{2+}$ are almost the same, that is, 57 pm for Mg$^{2+}$ and 60 pm for Zn$^{2+}$ in the case where the coordination numbers = 4 [8], this variation trend with changing $x$ implied that the chemical bond between (Mg,Zn) and N was more covalent than ionic because the covalent bond radius of Zn is slightly smaller than that of Mg. Note that a small amount of BN was observed at $2\theta$ = ~26.8°, which is indicated by vertical arrows in Fig. 1(a). It was very difficult to completely remove this BN impurity in some samples because a high pressure and a high temperature of 5.0 GPa and 1300 °C were applied to the mixed-powder disks tightly contacted with capsules made of BN in the high-pressure cells. However, this thin BN impurity located only around the outside wall of the resulting disk samples, that is, the BN impurity did not diffuse inside the bulk region of the samples. It should be noted that we observed that the phase stability in air became strong with increasing $x$, that is, the $x$ = 0 sample was unstable in air because it decomposed into an amorphous-like hydroxide $Ae$(OH)$_2$ ($Ae$ = Mg, Ca) just after air exposure, which is similar to the result reported on binary nitride precursors, Mg$_3$N$_2$ and Ca$_3$N$_2$ [9]. In contrast, the $x \geq 0.50$ samples were stable at room temperature in air. In addition, the sample purity strongly depended on the $x$ content. The volume fraction of the main phase and impurity concentration for each $x$ sample are summarized in Fig. 1(c). Phase purity of the $x = 0$ sample was the best (~90%) among all the $x$ contents. Addition of Zn decreased the volume fraction of the main phase to 50–60% for the $x = 0.29$–1 samples. Accordingly, the $Mg_3N_2$ impurity slightly increased up to ~20% at $x = 0.29$, and the metal Zn impurity drastically increased at $x \geq 0.69$.

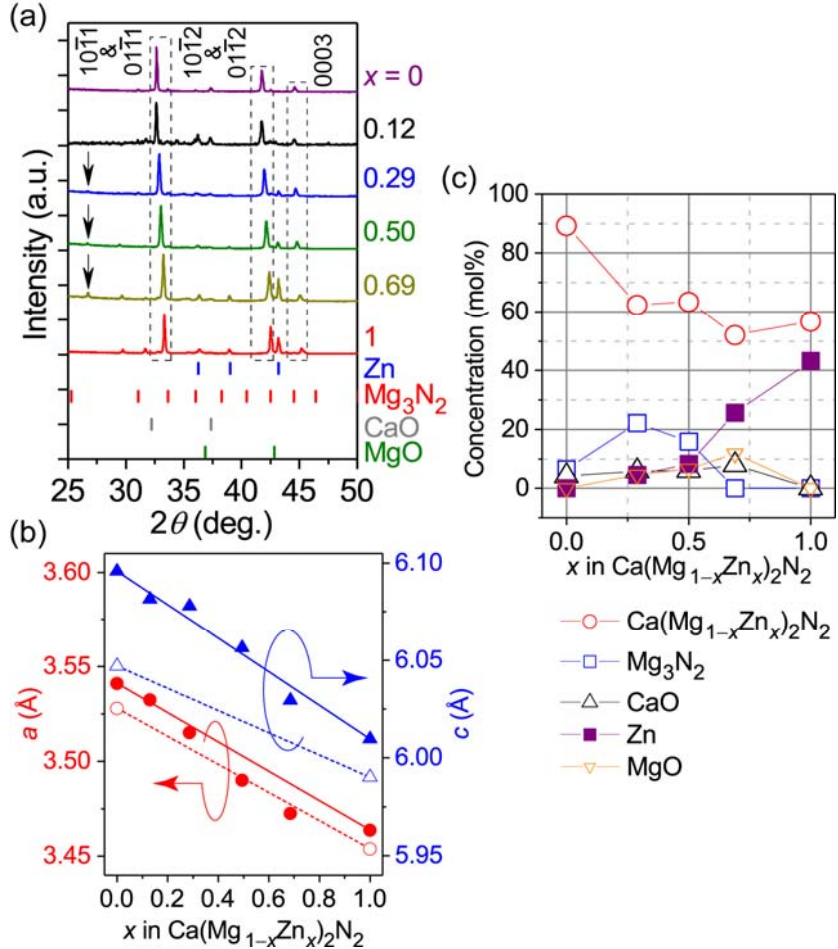

**Figure 1.** Structure, crystalline phase, and composition of $Ca(Mg_{1-x}Zn_x)_2N_2$ ($x = 0$–1) polycrystalline samples obtained. (a) XRD patterns. Three dotted squares denote areas of the diffraction peak positions, which are assigned as $10\bar{1}1$ & $01\bar{1}1$ ($2\theta = 32.6$–33.3°), $10\bar{1}2$ & $01\bar{1}2$ (41.7–42.5°), and 0003 (44.6–45.1°) diffractions from $Ca(Mg_{1-x}Zn_x)_2N_2$ main phase with a hexagonal structure of a trigonal system with $P\bar{3}m1$. Diffraction peak positions of Zn, $Mg_3N_2$, CaO, and MgO impurities are indicated at the bottom by vertical bars. A small amount of BN, which is the material of capsules in the high-pressure cells, was observed at the vertical arrows position (~26.8°). (b) Lattice parameters $a$ and $c$ as a function of $x$. Closed circles and triangles are $a$ and $c$ obtained in this study, respectively. Open circles and triangles are the calculated results with their structure relaxation reported in Ref. [6]. Straight lines are shown on the assumption of Vegard's law between both end members. (c) Volume fraction of the main phase and impurities as a function of $x$. All the data are taken from the results of the Rietveld analysis. The fitted raw results are provided in Fig. S1.

**Optical properties of Ca(Mg$_{1-x}$Zn$_x$)$_2$N$_2$ ($x$ = 0–1)**

Figure 2 shows the optical properties of polycrystalline Ca(Mg$_{1-x}$Zn$_x$)$_2$N$_2$ ($x$ = 0–1) at room temperature. With an increase in $x$, the absorption edge (nearly equal to the optical band gap energy in this case) continuously shifted to the lower energy side from ~3.2 eV to ~1.8 eV [see black slope lines for $(\alpha \cdot h\nu/s)^2$ versus $h\nu$ plots (data plotted by solid lines) in Fig. 2(a)]. This continuously controllable band gap agreed well with the calculated results [6] although all of them were slightly lower than the calculated values [see Fig. 2(b)]. Accordingly, band-to-band PL originating from each band gap was clearly observed even at room temperature [see open circles and vertical bars in Fig. 2(a)]. For the $x$ = 0 sample, an additional large and broad emission band that peaked at ~2.6 eV was observed. The decay curve of this emission band was decomposed into 2 components, which were a relatively short component of <~30 ns and a longer component of ~670 ns (see Fig. S2). Although the exact origins of these two lifetimes were unclear, we propose possible origins: the former could be related to a small amount (7 mol%) of the Mg$_3$N$_2$ impurity because the energy position is close to the band gap of Mg$_3$N$_2$ (~2.5 eV) [10] and/or a recombination originating from defects in Mg$_3$N$_2$ [11]. The latter could originate from defects such as donor–acceptor pairs in CaMg$_2$N$_2$ because the sample has a defective polycrystalline nature and the lifetime of ~670 ns is close to those of donor–acceptor pairs with a short donor–acceptor distance << 10 nm of conventional semiconductors [12]. For the $x$ = 0.29 sample, we confirmed that the emission lifetime of an emitting band that peaked at ~2.8 eV was ~7 ns (see Fig.

S3), which is comparable to the pulse width of the excitation laser we used, that is, the exact lifetime could not be estimated by this experiment. However, the information of a short lifetime was sufficient to assign it to a band-to-band transition. Therefore, we concluded that the ~2.8 eV PL band originates from a band-to-band emission for $x$ = 0.29. In addition, an emission band that peaked at a ~3.1 eV was observed. This PL originated from the BN impurity [13], because we also observed the same emission band from the $x$ = 0.50 and 0.69 samples when we set the observation wavelength region of the spectrometer around this region (data not shown) and a small amount of BN impurity was commonly observed in these 3 samples in their XRD patterns [see vertical arrows at $2\theta$ = ~26.8° in Fig. 1(a)]. Consequently, based on the above experimental optical properties at room temperature, we continuously and widely tuned the optical band gaps and emission bands from ~3.2 eV ($\lambda$ = ca. 390 nm, ultraviolet region) to ~1.8 eV (ca. 670 nm, red region) in complete solid-solution samples of $Ca(Mg_{1-x}Zn_x)_2N_2$ ($x$ = 0–1). In particular, it should be noted that the $x$ = 0.50 sample exhibited intense green light emission that peaked at ~506 nm (= ~2.45 eV) even at room temperature.

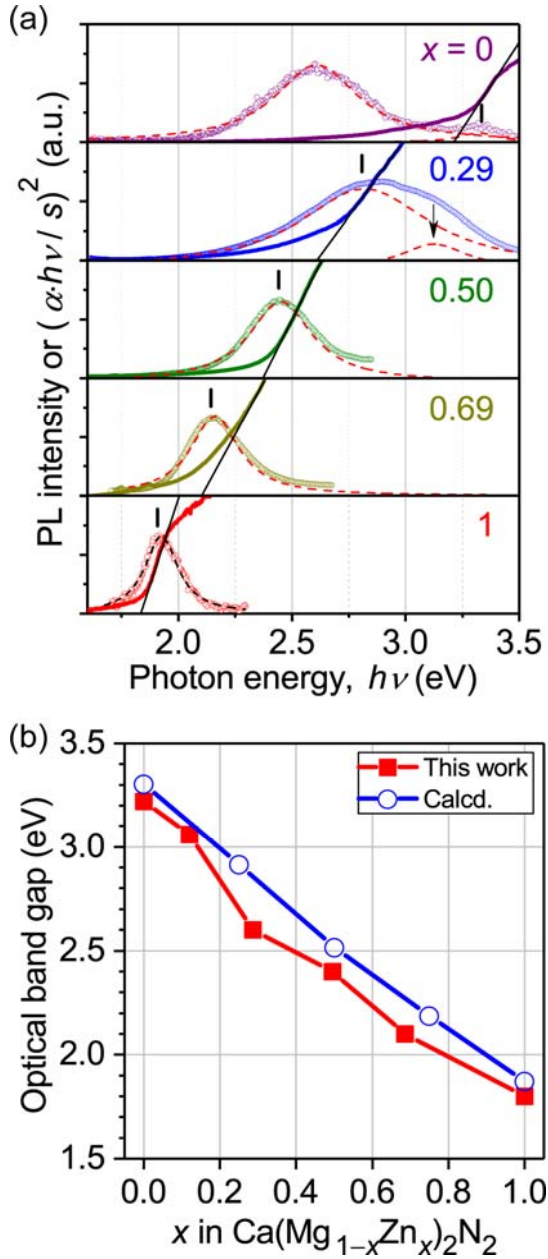

**Figure 2.** Optical properties of Ca(Mg$_{1-x}$Zn$_x$)$_2$N$_2$ ($x$ = 0–1) polycrystalline samples at room temperature. (a) PL (open circles) and $R$ (= $\alpha/s$, solid lines) spectra. The dotted lines are results of deconvolution of PL spectra. The PL peak positions arising from band-to-band transition are shown by vertical bars. The black slope lines for $(\alpha \cdot h\nu/s)^2$ versus $h\nu$ plots are used to estimate the absorption edges. A vertical arrow in the $x$ = 0.29 indicates an emission band from BN impurity. (b) The optical band gaps determined based on the plots of $(\alpha \cdot h\nu/s)^2$ versus $h\nu$ in (a) as a function of $x$. The calculated values (open circles) reported in Ref. [6] are shown for comparison.

In this study, we focused on light-emitting semiconductors in the green color region (i.e., band gap = 2.3−2.5 eV, $\lambda$ = 500−550 nm). Thus, we examined the PL properties of the $x$ = 0.50 sample in more detail because the PL band was located in the appropriate wavelength region for green emission. Figure 3(a) shows the temperature dependence of the PL spectra. We observed three emission bands, A–C. Band A originated from a small amount of a BN impurity [13], as discussed in the former section. Band B was assigned to the band-to-band transition because the emission peak position at 300 K was almost the same as that of the absorption edge observed in Fig. 2(a). Band C arose from deep-level defects and/or impurity phases because the energy deference between B and C was ~200 meV at 20 K, although the origin was unclear because the emission peak energy of band C did not depend on the measurement temperature and almost disappeared at ≥150 K. Because a similar temperature dependence has been reported for a Mg acceptor in GaN:Mg [14], band C could have similar origins, that is, point defects trapped at doped-impurity sites (e.g., band-to-impurity transitions). Owing to the band-to-band recombination, the energy of band B shifted to the lower energy side with increasing temperature from 2.49 eV at 20 K to 2.45 eV at 300 K [see Fig. 3(b)]. At 75–100 K, an anomalous PL peak shift from 2.48 eV to 2.49 eV (i.e., a blue shift with increasing temperature) was observed. The possible origin of this anomalous shift was as a result of a transition from bound excitons to free excitons, as observed in room-temperature-exciton (i.e., electron − hole pairs stable at room temperature) materials, such as ZnO [15] and LaCuOS [16]; whereas details of its origin should be discussed using more high-resolution and continuous wave-excited PL measurements on higher quality samples, such as single crystals and epitaxial films. The PL intensity was decreased by approximately one order of magnitude from 20 K to 300 K, owing to temperature deactivation [see Fig. 3(b)]. However, the emission observed in the $x$ = 0.50 sample shown in Fig. 2(a) was also clearly observed at room temperature. Next, we examined the emission lifetime of bands B and C. The estimated lifetimes at 20 K were ~2 ns for band B and ~250 ns for band C (see Fig. S4). Because the pulse width of the excitation Nd:YAG we used was ~7 ns, the lifetime of band B could not be exactly estimated. This result suggested that the emission band B originated from bound/free excitons because the lifetime was sufficiently shorter than several nanoseconds. The relatively fast lifetime of ~250 ns for band C was roughly consistent with the above proposed origin, that is, point defects trapped at doped-impurity sites because the lifetime of Mg acceptor in GaN:Mg has been reported to be on the subnanosecond order [17]. However, these above assignments are tentative and speculated possibilities.

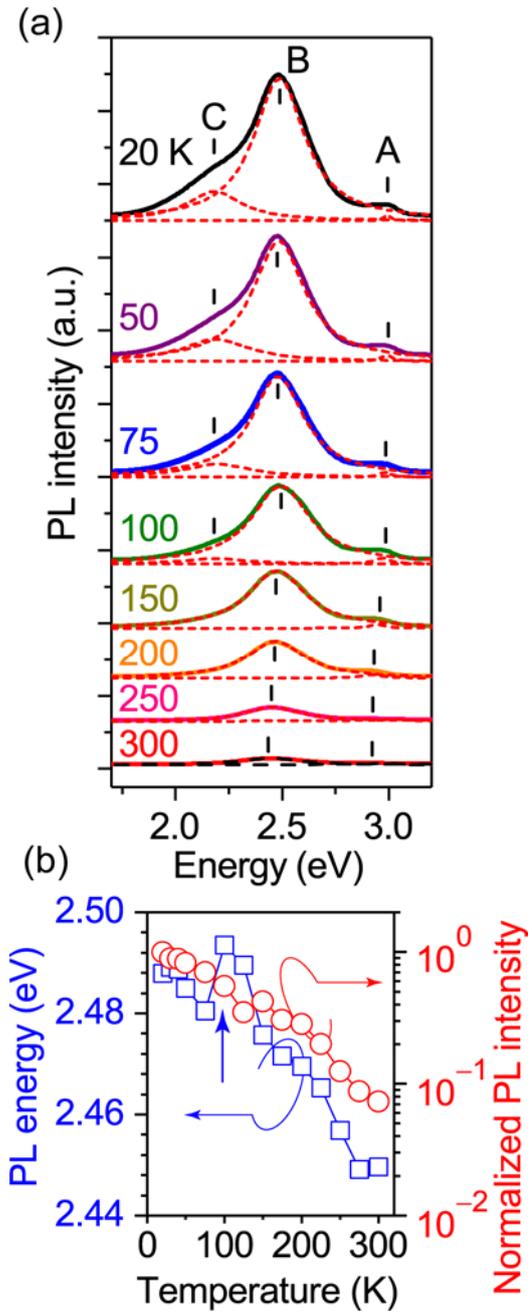

**Figure 3.** (a) Temperature dependence of PL spectra of Ca(Mg$_{1-x}$Zn$_x$)$_2$N$_2$ ($x$ = 0.50) polycrystalline sample. The dotted lines are deconvolution results of each PL spectrum. Note that a spectrum at 300 K is the same as that for $x$ = 0.50 in Fig. 2(a). (b) PL peak energy (squares, left axis) and normalized PL intensity (circles, right axis) of band B originating from band-to-band transition as a function of measurement temperature. The vertical arrow indicates the position of an anomalous blue shift between 75–100 K.

**Electronic transport properties of Na 2%-doped $(Ca_{1-y}Na_y)(Mg_{1-x}Zn_x)_2N_2$ ($x = 0.67$ and $y = 0.02$)**

Although Na-undoped $(Ca_{1-y}Na_y)(Mg_{1-x}Zn_x)_2N_2$ ($x = 0$–$1$ and $y = 0$) samples were all too highly resistive to measure the electrical conductivity (much higher than $10^6$ Ω cm), 2% Na-doping exhibited measurable conductivity. Figure 4(a) shows an enlargement of the strongest $10\bar{1}1$ & $0\bar{1}11$ peak region ($2\theta$ = ca. 33°) in the XRD pattern of the Na 2%-doped polycrystalline sample. The Rietveld analysis result is shown in Fig. S5. The main phase purity was low, 25%, while those of Zn and $Mg_3N_2$ impurities that might possibly exhibit electrical conduction were also small, 8% and 15%, respectively. Thus, it was considered that such small amounts of possibly electrical conductive impurities should not dominate the electrical conduction of the sample. Irrespective of almost the same ionic radii of $Ca^{2+}$ and $Na^+$, the XRD peak slightly shifted to the higher $2\theta$ side by Na doping, which indicated its lattice shrinkage. This variation trend of lattice parameters was attributed to the participation of a covalent nature in the bonding between (Ca,Na) and N, as was also observed in Fig. 1(b). Although the undoped sample with $x \geq 0.5$ was chemically stable, the Na-doped sample with $x \geq 0.5$ was quite sensitive to air exposure. The temperature dependence of $\rho$ exhibited a thermally activated-type behavior, which implied that metallic zinc did not contribute to the electronic conduction. The inset of Fig. 4(b) is the result of a thermoelectric power measurement at room temperature. These data indicate p-type electrical conduction with a positive Seebeck coefficient of +2.3 µV/K. However, the activation energy estimated using an Arrhenius plot (see Fig. S6) was ~330 meV, which suggested that relatively deep acceptor levels formed within the energy band gap upon Na doping. We also tried n-type carrier doping by adding La or Y to the Ca site, but no enhancement of electrical conduction was observed. These results indicated that we could successfully obtain a p-type semiconducting sample with an appropriate band gap near green emission by selecting proper $x$ and Na doping amounts.

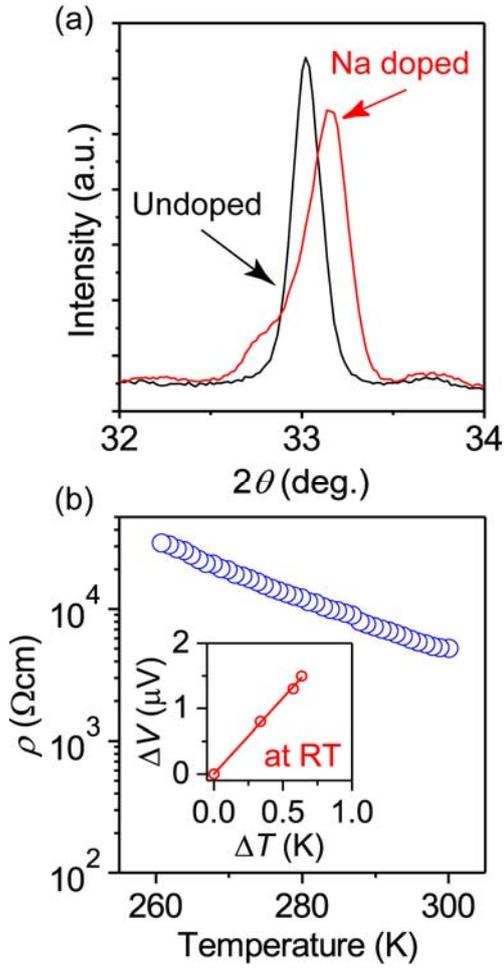

**Figure 4.** Structure and electronic transport properties of the $(Ca_{1-y}Na_y)(Mg_{1-x}Zn_x)_2N_2$ ($x = 0.67$ and $y = 0.02$) polycrystalline sample. (a) XRD diffraction pattern around $10\bar{1}1$ & $01\bar{1}1$ (red pattern), which is the strongest peak of the main phase. The peak of undoped sample (black pattern) is plotted for comparison. (b) Temperature dependence of electrical resistivity ($\rho$). The inset shows thermoelectric power ($\Delta V$) as a function of difference in temperature ($\Delta T$) at room temperature (RT), revealing successful carrier doping and its p-type electrical conductivity at room temperature.

In summary, we synthesized $Ca(Mg_{1-x}Zn_x)_2N_2$ ($x = 0–1$) by conventional solid-state reaction for $x = 0$ and $0.12$ and high-pressure synthesis at 5 GPa for $0.12 < x \leq 1$ in $Ca(Mg_{1-x}Zn_x)_2N_2$. It was experimentally validated that the optical band gaps can be tuned continuously from ~3.2 eV to ~1.8 eV. Owing to the direct transition-type band structure, ultraviolet–red band-to-band photoluminescence was clearly observed in the entire $x = 0$–1 region even at room temperature. p-type electrical conduction was also demonstrated by 2% Na doping at the Ca site. Especially, the sample with $x = 0.50$ exhibited an intense green band-to-band emission, which is strongly demanded in III-V group light-emitting semiconductors.

**Experimental Methods**

Three polycrystalline binary nitrides, calcium nitride ($Ca_3N_2$), magnesium nitride ($Mg_3N_2$), and zinc nitride ($Zn_3N_2$), were used as starting precursors for undoped ($Ca_{1-y}Na_y$)($Mg_{1-x}Zn_x$)$_2N_2$ ($x = 0$–1, $y = 0$, i.e., not hole carrier-doped) polycrystalline samples. $Ca_3N_2$ powder was synthesized by heating Ca metal (purity: 99.99%), which was put on a rolled Fe foil in a silica-glass tube, at 900 °C for 10 h under a $N_2$ gas flow atmosphere (G1 grade, flow rate of $N_2$ gas = 100 mL/min.). For $Mg_3N_2$ powder, Mg metal (purity: 99.5%) was reacted with $N_2$ gas in almost the same manner as that of the $Ca_3N_2$ precursor. All these processes, including heating, were performed in an Ar-filled glove box (oxygen concentration ≤ 0.0 ppm, dew point < −100 °C) without exposure to air because both the binary nitrides are sensitive to air; that is, the silica-glass tube furnace was directly connected to the glove box. Such a direct nitridization process of alkali earth metals under $N_2$ atmosphere is similar with previous reports [18, 19]. The $Zn_3N_2$ powder we used was a commercially available product (purity: 99%). Colors of the precursor powders were rubric black for $Ca_3N_2$, dark chrome-yellow for $Mg_3N_2$, and gray for $Zn_3N_2$, which were roughly consistent with their reported optical band gaps of ~1.5 eV for $Ca_3N_2$ [20], ~2.5 eV for $Mg_3N_2$ [10], and ~0.85 eV for $Zn_3N_2$ [21]. $Ca_3N_2$ and $Mg_3N_2$ powders were synthesized just before mixing and heating experiments for syntheses of $Ca(Mg_{1-x}Zn_x)_2N_2$ ($x = 0$–1) and were completely used within two days as it was difficult to store them without degradation even in the glove box for a long time, such as ~1 week.

To synthesize polycrystalline $Ca(Mg_{1-x}Zn_x)_2N_2$ ($x = 0$ and 0.12), a conventional solid-state reaction was employed. Stoichiometric mixtures of $Ca_3N_2$, $Mg_3N_2$, and Zn$_3$N$_2$ precursor powders with the molar ratios of Ca$_3$N$_2$:Mg$_3$N$_2$:Zn$_3$N$_2$ = 1:2(1−$x$):2$x$ were pressed into disks (size: diameter of ~6 mm and height of ~7 mm) in the glove box, which were sealed in Ar-filled stainless tubes, followed by a heating process at 1050 °C for 24 h, as has previously been reported for CaMg$_2$N$_2$ [7].

In contrast, to synthesize Zn-substituted nitrides with higher $x$, Ca(Mg$_{1−x}$Zn$_x$)$_2$N$_2$ (0.12 < $x$ ≤ 1) solid-solutions, we used a belt-type high-pressure apparatus [22] to realize a high nitrogen chemical potential during the chemical reaction between precursors in the high-pressure cells. Stoichiometric mixtures of each binary precursor were pressed into disks, which were put into BN capsules, followed by setting up in high-pressure cells made of NaCl (+10 wt% ZrO$_2$) with a graphite heater. We also examined Na doping at the Ca site for hole-carrier doping because the ionic radius of Na is close to that of Ca (100 pm for Ca$^{2+}$ and 102 pm for Na$^+$ for the case where their coordination numbers = 6) [8], whereas the atomic/covalent bond radius of Na is slightly smaller than that of Ca. A commercially available NaN$_3$ reagent powder (99.5%) was used as a sodium source. As-received NaN$_3$ was thermally annealed in a vacuum of ~10$^{−2}$ Pa at 250 °C for 6 h to purify before mixing. We mixed the precursor powders so as to be (Ca$_{1−y}$Na$_y$)(Mg$_{1−x}$Zn$_x$)$_2$N$_2$ ($x$ = 0.67 and $y$ = 0.02).

The pressure and heating temperature of the high-pressure cells, including BN capsules and pressed disks in the glove box for the synthesis of undoped and Na-doped Ca(Mg$_{1−x}$Zn$_x$)$_2$N$_2$ (0.12 < $x$ ≤ 1), were commonly increased to 5.0 GPa and 1300 °C, respectively, and they were then kept for 30 minutes. The final size of the high-pressure-synthesized samples was a diameter of ~5.5 mm and thickness of ~2 mm (after polishing). More details of the experimental procedure and apparatus of the high-pressure synthesis is reported in Ref. [6].

Crystal structures of the samples, such as crystalline phases and lattice parameters, were determined by powder X-ray diffraction (XRD) with the Bragg-Brentano geometry with a Cu K$\alpha$ radiation source at room temperature, where an Ar-filled O-ring-sealed sample holder was used because the Mg-rich samples ($x$ < 0.5) were sensitive to air. The lattice parameters of the main phase were determined by the Pawley method using the TOPAS ver. 4.2 program (Karlsruhe, Germany: Bruker AXS GmbH).

Rietveld analysis was performed to estimate the volume fraction of each crystalline phase observed and $x$ in Ca(Mg$_{1-x}$Zn$_x$)$_2$N$_2$ [i.e., crystallographic site occupancy of the (Mg,Zn) site] using the TOPAS program after initially determining the lattice parameters by the Pawley method because it was difficult to exactly distinguish the main phase of Ca(Mg$_{1-x}$Zn$_x$)$_2$N$_2$ from some impurity phases, such as Zn and Mg$_3$N$_2$, by conventional spectroscopic quantitative techniques, such as X-ray fluorescence spectroscopy and electron probe microanalysis.

Diffuse reflectance ($R$) spectra were measured at room temperature with a spectrophotometer in the wavelength ($\lambda$) region of 200–2400 nm. The obtained $R$ spectra were converted using the Kubelka−Munk function $(1–R)^2/(2R) = \alpha/s$ [23], where $\alpha$ and $s$ denote the optical absorption coefficient and scattering factor, respectively, to obtain the quasi-optical absorption spectra. The optical band gap of each sample was estimated from plots of $(\alpha \cdot h\nu/s)^2$ versus $h\nu$, where $h$ and $\nu$ are the Planck constant and frequency, respectively, because the band structures of both end members (i.e., $x = 0$ and 1) have been reported to be of the direct transition type [6]. Photoluminescence (PL) spectra and emission lifetimes at 20–300 K were attained with a CCD camera and a spectrometer under excitation by $3\omega$ of a Nd:YAG laser ($\lambda = 355$ nm, pulse width = ~7 ns); whereas only for the CaMg$_2$N$_2$ (i.e., $x = 1$) sample, $4\omega$ of the Nd:YAG ($\lambda = 266$ nm) was used because of the widest bandgap in Ca(Mg$_{1-x}$Zn$_x$)$_2$N$_2$ ($x = 0$–1).

Electronic transport properties were measured only for the (Ca$_{1-y}$Na$_y$)(Mg$_{1-x}$Zn$_x$)$_2$N$_2$ ($x = 0.67$ and $y = 0.02$) sample because the electrical resistivity of the other undoped samples (i.e., $x = 0$–1 and $y = 0$) was too high to measure their transport properties. We initially confirmed the electrical conduction with a conventional electric tester in the glove box because the Na-doped sample was air sensitive. The temperature dependence of the electrical resistivity, and thermoelectric power ($\Delta V$) as a function of difference in temperature ($\Delta T$) at room temperature (i.e., the Seebeck effect) were measured with a physical property measurement system. For these electrical measurements, we transferred the Na-doped sample from the glove box to the measuring system within ~1 min together with surface protection using commercially available Apiezon grease to disturb the degradation.

**Acknowledgments**

This work was supported by JST CREST Grant No. JPMJCR17J2, Japan. H. Hosono was supported by the Ministry of Education, Culture, Sports, Science, and Technology (MEXT) through the Element Strategy Initiative to Form Core Research Center. H. Hiramatsu was also supported by the Japan Society for the Promotion of Science (JSPS) through Grants-in-Aid for Scientific Research (A) and (B) (Grant Nos. 17H01318 and 18H01700), and Support for Tokyotech Advanced Research (STAR).

**Supporting Information**

Supporting information is available at ACS web site for XRD patterns and their Rietveld analysis results, and emission decay curves.

Supporting Information for "Tunable light-emission through the range 1.8–3.2 eV and p-type conductivity at room temperature for nitride semiconductors, Ca(Mg$_{1-x}$Zn$_x$)$_2$N$_2$ ($x = 0 - 1$)"

Masatake Tsuji,[1] Hidenori Hiramatsu,[1,2] and Hideo Hosono[1,2]

1: Laboratory for Materials and Structures, Institute of Innovative Research, Tokyo Institute of Technology, Mailbox R3-3, 4259 Nagatsuta-cho, Midori-ku, Yokohama 226-8503, Japan
2: Materials Research Center for Element Strategy, Tokyo Institute of Technology, Mailbox SE-1, 4259 Nagatsuta-cho, Midori-ku, Yokohama 226-8503, Japan

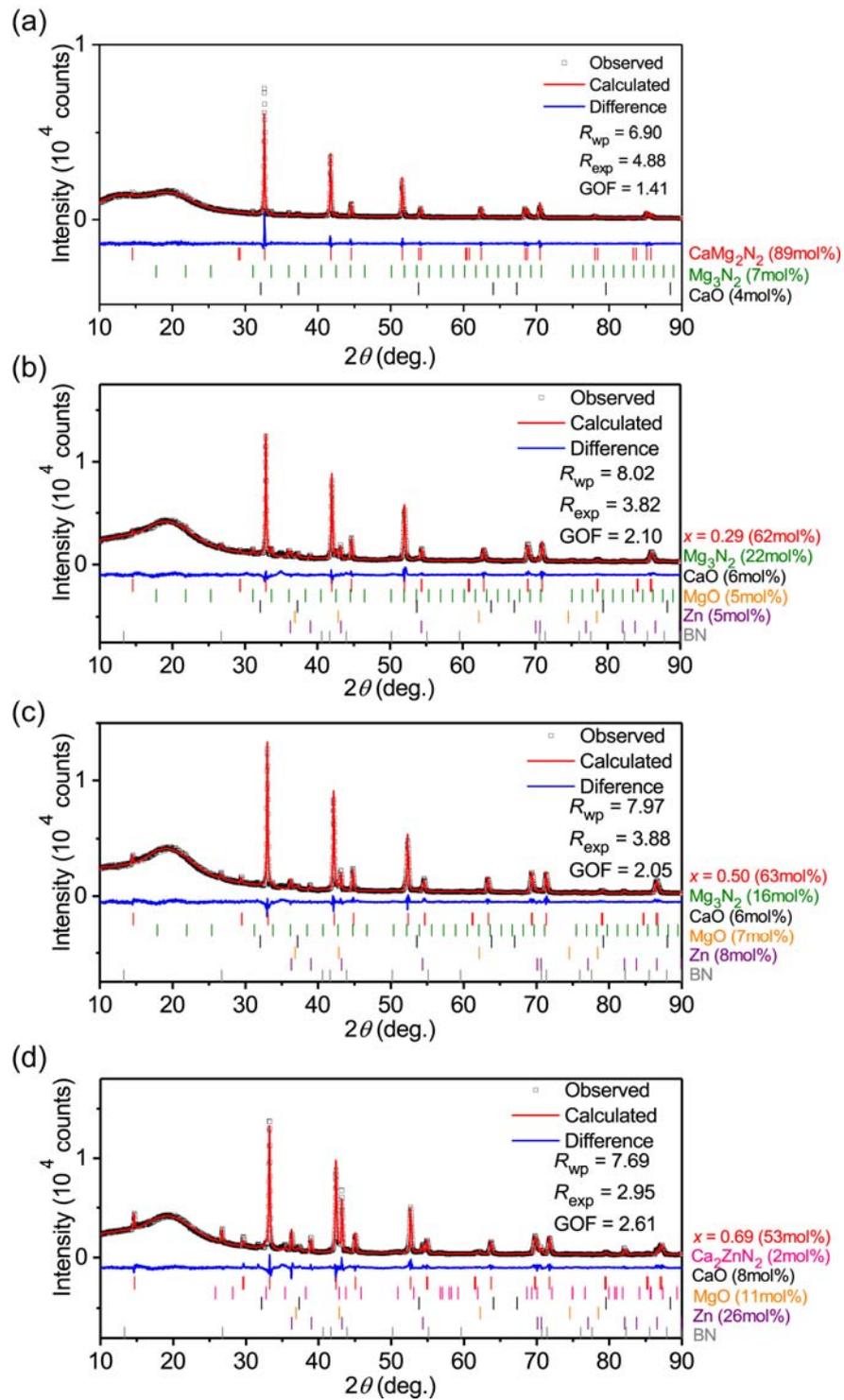

**Figure S1.** Results of Rietveld analysis for $Ca(Mg_{1-x}Zn_x)_2N_2$ [(a) $x = 0$, (b) 0.29, (c) 0.50, and (d) 0.69] polycrystalline samples. Broad halo patterns observed at $2\theta$ = ca. 20° originate from the Ar-filled O-ring-sealed sample holder. The result of the $x = 1$ sample (i.e., $CaZn_2N_2$) can be found in Ref. [Y. Hinuma et al., Nat. Commun. 7, 11962 (2016).].

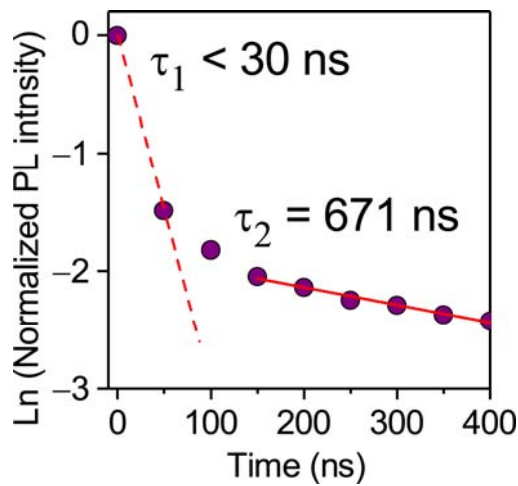

**Figure S2.** Time decay of PL intensity at room temperature of an emission band that peaked at ~2.6 eV observed in CaMg$_2$N$_2$ (i.e., $x = 0$) polycrystalline sample [Fig. 3 (a) in the main text]. Closed circles and the two straight lines denote raw data and fitted results based on each single decay model, respectively.

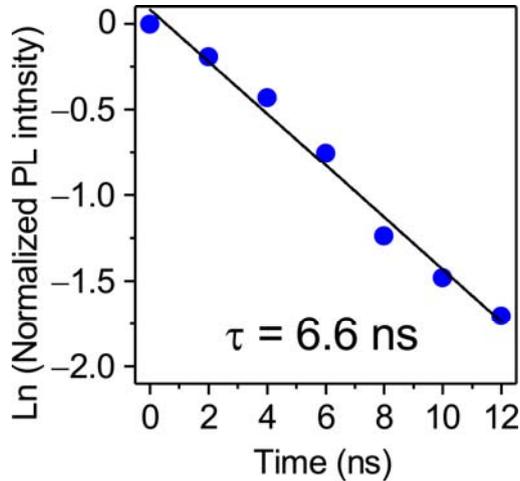

**Figure S3.** Time decay of PL intensity at room temperature of an emission band that peaked at ~2.8 eV, which originates from band-to-band transition observed in Ca(Mg$_{1-x}$Zn$_x$)$_2$N$_2$ ($x = 0.29$) polycrystalline sample [Fig. 3(a) in the main text]. Closed circles and a straight line denote raw data and fitted result based on a single decay model, respectively.

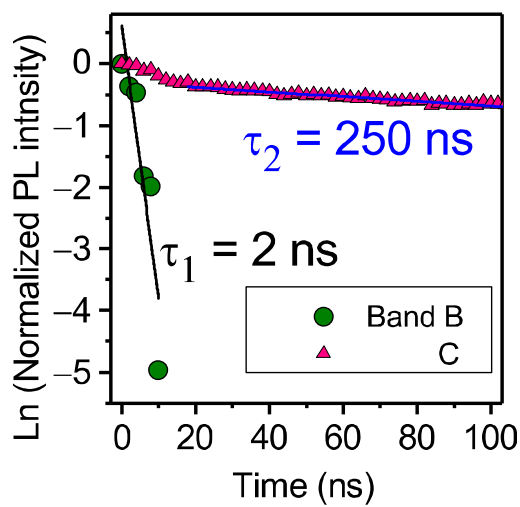

**Figure S4.** Time decay of PL intensity of emission bands B and C for Ca(Mg$_{1-x}$Zn$_x$)$_2$N$_2$ ($x$ = 0.5) polycrystalline sample at 20 K [Fig. 3(a) in the main text]. Closed symbols and straight lines denote raw data and fitted results based on a single decay model, respectively.

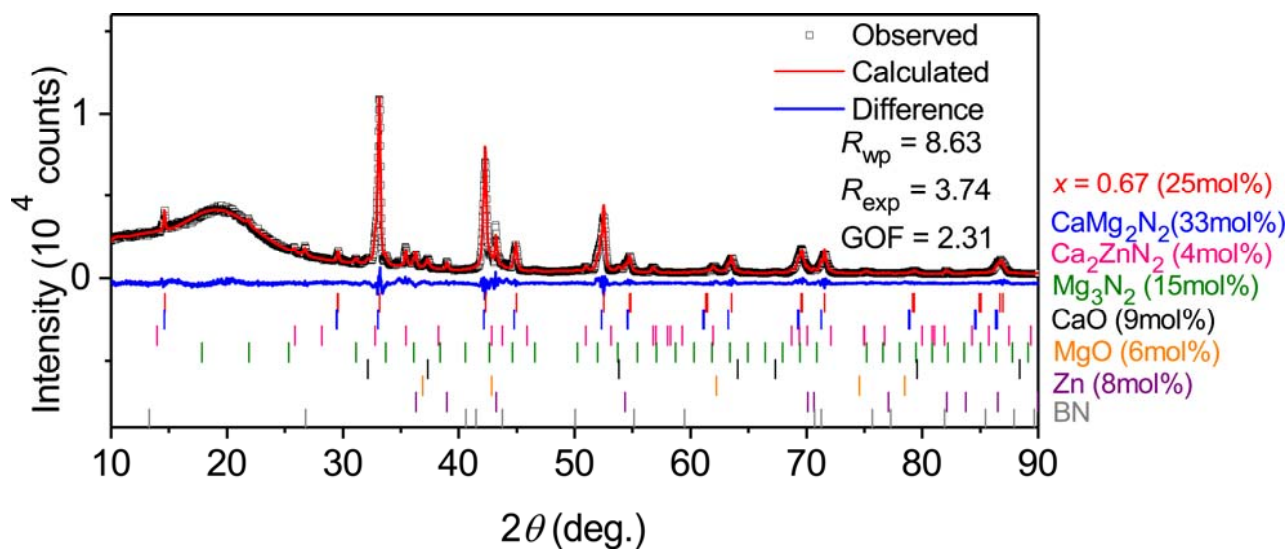

**Figure S5.** Result of Rietveld analysis for (Ca$_{1-y}$Na$_y$)(Mg$_{1-x}$Zn$_x$)$_2$N$_2$ ($x$ = 0.67, $y$ = 0.02) polycrystalline sample. A broad halo pattern observed at $2\theta$ = ca. 20° originates from the Ar-filled O-ring-sealed sample holder.

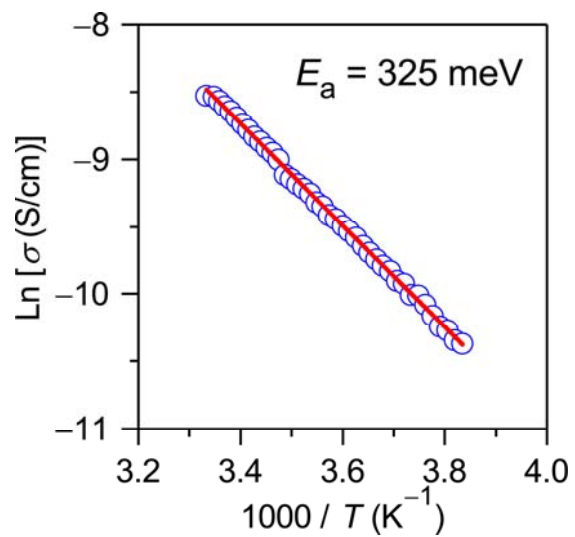

**Figure S6.** Arrhenius plot for $(Ca_{1-y}Na_y)(Mg_{1-x}Zn_x)_2N_2$ ($x$ = 0.67, $y$ = 0.02) to estimate activation energy $E_a$. $\sigma$ denotes electrical conductivity with a relation of $\sigma = 1/\rho$.